\documentclass[nohyper,12pt,letterpaper]{JHEP}
\usepackage[dvips]{epsfig}
\usepackage{epsf,amsfonts,amssymb}

\newcommand{\eqn}[1]{(\ref{#1})}
\newcommand{\be}{\begin{equation}}
\newcommand{\ee}{\end{equation}}
\newcommand{\ben}{\begin{displaymath}}
\newcommand{\een}{\end{displaymath}}
\newcommand{\bea}{\begin{eqnarray}}
\newcommand{\eea}{\end{eqnarray}}
\newcommand{\bean}{\begin{eqnarray*}}
\newcommand{\eean}{\end{eqnarray*}}
\newcommand{\nn}{\nonumber \\}
\newcommand{\ba}{\begin{array}}
\newcommand{\ea}{\end{array}}
\newcommand{\bi}{\begin{itemize}}
\newcommand{\ei}{\end{itemize}}

\def\a {\alpha}
\def\b {\beta}
\def\g {\gamma}

\def\d {\delta}

\def\S {\Sigma}
\def\e {\epsilon}

\def\vp {\varphi}

\renewcommand{\O}{\Omega}

\newcommand{\cala}{\mbox{${\cal A}$}}

\newcommand{\calc}{\mbox{${\cal C}$}}
\newcommand{\cald}{\mbox{${\cal D}$}}

\newcommand{\calh}{\mbox{${\cal H}$}}

\newcommand{\call}{\mbox{${\cal L}$}}

\newcommand{\caln}{\mbox{${\cal N}$}}
\newcommand{\calo}{\mbox{${\cal O}$}}
\newcommand{\calp}{\mbox{${\cal P}$}}

\newcommand{\calt}{\mbox{${\cal T}$}}

\newcommand{\calz}{\mbox{${\cal Z}$}}

\newcommand{\bbe}[1]{{\mathbb E}^{#1}}

\newcommand{\bbi}[1]{{\mathbb I}_{#1}}
\newcommand{\bbo}[1]{{\mathbb O}_{#1}}

\newcommand{\bra}[1]{\mbox{$\langle #1 |$}}
\newcommand{\ket}[1]{\mbox{$| #1 \rangle$}}
\newcommand{\braket}[2]{\mbox{$\langle #1  | #2 \rangle$}}

\newcommand{\Gn}{\Gamma_\natural}

\newcommand{\pa}{\partial}
\newcommand{\fc}{\frac}
\newcommand{\w}{\wedge}
\newcommand{\trace}{\mbox{Tr}}
\newcommand{\sac}{\, , \qquad}
\newcommand{\eg}{{\it e.g. }}

\newcommand{\sgn}{\mbox{sgn}}

\newcommand{\ik}{{\it k}}
\newcommand{\ione}{{\it 1}}
\newcommand{\itwo}{{\it 2}}
\newcommand{\ph}{\hat{P}}
\newcommand{\xh}{\hat{X}}
\newcommand{\ve}{\varepsilon}
\newcommand{\tr}{\mbox{Tr}}

\newcommand{\sect}[1]{\setcounter{equation}{0}\section{#1}}
\renewcommand{\theequation}{\arabic{section}.\arabic{equation}}

\title{Tachyons, Supertubes\\ \hspace{4cm} and Brane/Anti-Brane Systems}

\author{David Mateos, Selena Ng and Paul K.\ Townsend \\
   Department of Applied Mathematics and Theoretical Physics\\
   Centre for Mathematical Sciences \\
   Wilberforce Road, Cambridge CB3 0WA, United Kingdom \\
E-mail: \email{D.Mateos, S.K.L.Ng, P.K.Townsend@damtp.cam.ac.uk}}

\abstract{We find supertubes with {\it arbitrary} (and not necessarily
planar) cross-section; the stability against the D2-brane tension 
is due to a compensation by the local momentum generated by 
Born-Infeld fields. Stability against long-range supergravity forces 
is also established. We find the corresponding solutions of the 
$N=\infty$ M(atrix) model. The supersymmetric $D2$/${\overline{D2}}$
system is a special case of the general supertube, 
and we show that there are no open-string tachyons in this system 
via a computation of the open-string one-loop 
vacuum energy.}


\preprint{DAMTP-2001-106 \\ \tt{hep-th/0112054}}

\begin{document}


\sect{Introduction}

A collection of branes may (under some circumstances) 
find it energetically favourable to expand to form a brane of higher
dimension \cite{roberto,myers}. This `brane expansion' 
plays a fundamental role in a number of phenomena, such as the 
behaviour of gravitons at high energies in certain backgrounds
\cite{giant} or the string realization of the vacuum structure of 
$\caln$=1 four-dimensional gauge theories \cite{ps}. The presence 
of the new brane `created' by the expansion is implied by the 
fact that the expanded system couples locally to higher-rank 
gauge fields under which the original
constituents are neutral\footnote{This feature is absent from 
a phenomenon such as the enhan\c{c}on \cite{enhancon}  which some
authors also refer to as `brane expansion'.}. For a large number of
constituent branes there may then be an effective description in terms
of the higher-dimensional brane in which the original branes have
become fluxes of various types. Since no {\it net} higher-dimensional 
brane charge is created by the process of expansion, at least 
one of the dimensions of the higher-dimensional brane must be 
compact and homologically trivial.

Although the phenomenon of brane expansion was originally discovered
in the context of supersymmetric theories, early examples of
`expanded-brane' configurations were not themselves supersymmetric,
and hence unlikely to be stable. Supersymmetric expanded brane 
configurations in certain backgrounds have been found\footnote{
For example (but not exclusively) in $AdS$ backgrounds
\cite{giant,ps}. We thank Iosif Bena for correspondence on this point.} 
but the presence of non-vanishing Ramond-Ramond fields in all of 
them makes any conventional string theory analysis difficult. 
At present, the only example of a
supersymmetric expanded brane configuration in a vacuum background 
is the D2-brane `supertube' of \cite{MT01}, which was provided with a
Matrix Theory interpretation in \cite{BL01}. Supertubes
are collections of type IIA fundamental strings and D0-branes which 
have been expanded, in the IIA Minkowski
vacuum, to tubular 1/4-supersymmetric D2-branes by the addition of
angular momentum\footnote{{\it Non}-supersymmetric brane  expansion by
angular momentum in Minkowski space was considered previously in
\cite{savvidy}.}. 

The simplest potential instability of any expanded brane configuration is
that caused by brane tension, which tends to force the system to contract.
The presence or absence of this instability is captured by the low-energy
effective action for the brane in question, \eg the
Dirac-Born-Infeld (DBI) action for D-branes. The D2-brane DBI action is
what was used in \cite{MT01} to find the D2-brane supertube (with circular
cross-section) and to establish its supersymmetry.  Although the original
supertube was assumed to have a circular cross-section, it was shown in
\cite{BK01}, in the Matrix Theory context, that an elliptical cross-section
is also possible. One purpose of this paper is to generalize the results
of \cite{MT01} to show that 1/4-supersymmetric supertubes may have 
as a cross-section a completely {\it arbitrary} curve in $\bbe{8}$. 
The stability of the original supertube was
attributed to the angular momentum generated by the Born-Infeld (BI) 
fields.  Although this is still true, the stability of the general
supertube configurations is less easily understood; as we shall see the
D2-brane tension is not isotropic and the supertube behaves in some
respects like a {\it tensionless} brane. We also find these general
supertube configurations as solutions of the light-front gauge-fixed
eleven-dimensional supermembrane equations. As this is the $N=\infty$ limit of
the Matrix  model, we thus make contact with the Matrix Theory approach
of \cite{BL01,BK01}. 

A second potential instability is that associated to the long-range forces
between different regions of an expanded brane (or between different
expanded branes) due to the exchange of massless particles. In the
case of the D2-brane supertube these are particles in the
closed IIA superstring spectrum. Whether or not this instability 
occurs may be determined by considering the D2-brane supertube 
in the context of the effective IIA supergravity theory. 
For circular supertubes this aspect was considered in \cite{EMT01}, 
where a 1/4-supersymmetric supergravity solution for a general
multi-supertube system was constructed and stability against
supergravity forces confirmed\footnote{Multi-supertube systems 
have been considered in the context of Matrix Theory in
\cite{BL01}.}. Here we shall generalize these results to supertubes
with arbitrary cross-section by explicitly exhibiting the
corresponding supergravity solutions.

In the case of D-branes (at least) there is a third more dramatic 
and purely `stringy' potential instability of an expanded brane
configuration arising from the fact that opposite sides (along one 
of the compact directions) of the (higher-dimensional) D-brane 
behave locally as a brane/anti-brane pair. For sufficiently small
separation there are tachyons in the open strings between a D-brane
and an anti-D-brane, so a sufficiently compact expanded D-brane
configuration is potentially unstable against tachyon condensation 
\cite{sen}. One purpose of this paper is to examine this issue for 
the D2-brane supertube. One would certainly expect it to be
stable given that it preserves supersymmetry, but supersymmetry 
has been established only within the context of the effective DBI
action and the effective supergravity theory. Thus the preservation of
supersymmetry could be an artifact of these effective theories that is
not reproduced within the full string theory; the stability of the
supertube in the full IIA string theory is therefore not guaranteed 
{\it a priori}. 

The main obstacle to a full string theory analysis of the supertube
is that, as for any other expanded brane configuration, the D-brane 
surface must be (extrinsically) curved, so even though we have the
advantage of a vacuum spacetime background the quantization of open
strings attached to the D2-brane supertube is a difficult problem.
However, it seems reasonable to suppose that any tachyons in the
supertube system would also appear in the case in which 
the D2-brane is locally approximated by a flat `tangent-brane'.
In fact, because the cross-section of the supertube is {\it arbitrary}
we may take a limiting case in which the tube becomes a pair of
D2-branes which intersect at an arbitrary angle $\phi$;
the particular case $\phi=\pi$ corresponds to a parallel D2/anti-D2
pair, discussed in \cite{BK01} in the Matrix model
context. Another purpose of this paper is to show that there is no
tachyon in this system for any angle.
We establish this by computing
the one-loop vacuum energy of the open strings stretched between 
the D2-branes.

\sect{Supertubes with Arbitrary Cross-sections}
\label{DBI}


In the original paper on supertubes \cite{MT01} only circular
cross-sections were considered. This was later generalized in
the context of Matrix theory \cite{BK01} to elliptical
cross-sections. The purpose of this section is to show that any
tubular D2-brane with an arbitrary cross-section (not necessarily
contained in a 2-plane nor closed) carrying the appropriate string 
and D0-brane charges is also 1/4-supersymmetric. 
This point is not immediately obvious from the approach in \cite{BK01}
and it was missed in the DBI analysis of \cite{MT01}.

\subsection{Supersymmetry}

Consider a D2-brane with worldvolume coordinates $\xi^a=\{t,z,\sigma\}$
in the type IIA Minkowski vacuum. We write the spacetime metric as
\be
ds_{\it 10}^2 = - dT^2 + dZ^2 + d\vec{Y} \cdot d\vec{Y} \,,
\label{st-metric}
\ee
where $\vec{Y}=\{Y^i\}$ are Cartesian coordinates on $\bbe{8}$, and set
\be
T=t \sac Z=z \sac \vec{Y}=\vec{y}(\sigma) \,.
\ee 
This describes a static tubular D2-brane whose axis is aligned
with the $Z$-direction and whose cross-section is an arbitrary curve 
$\vec{y}(\sigma)$ in $\bbe{8}$ (see Figure \ref{tube}).
Although the term `tubular' is strictly appropriate only when the 
curve is closed, we will use it to refer to any of these configurations. 

\FIGURE{
\centerline
{\epsfig{file=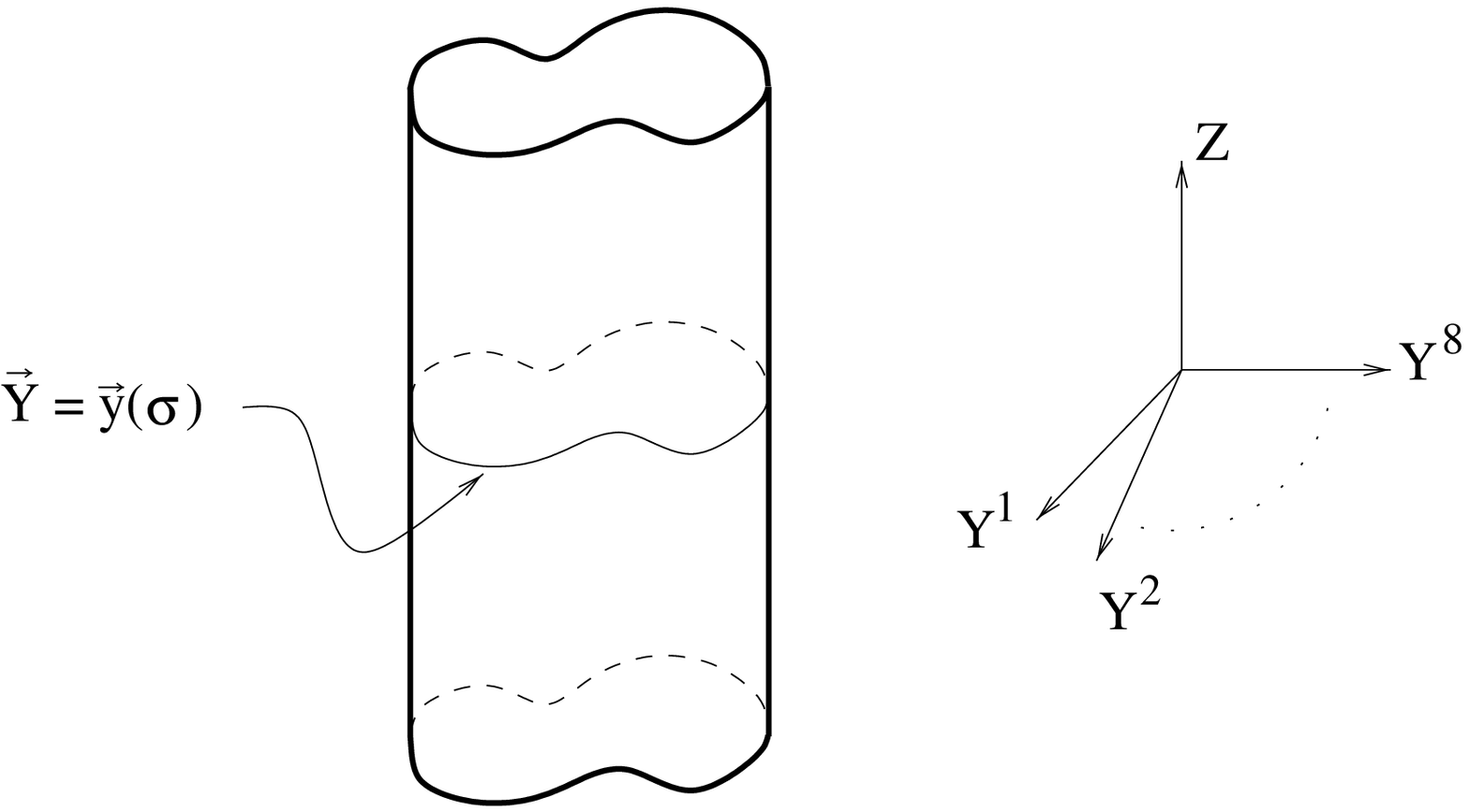, height=7cm}}
\caption{Supertube with arbitrary cross-section 
$\vec{Y} = \vec{y}(\sigma)$ in $\bbe{8}$.}
\label{tube}
}

We will also allow for time- and $z$-independent electric 
(in the $Z$-direction) and magnetic fields, so the BI field-strength on
the D2-brane is  
\be
F = E \, dt \wedge dz + B(\sigma) \, dz \wedge d\sigma \,.
\ee
As we will see in more detail below, this corresponds to having string
charge in the $Z$-direction and D0-brane charge dissolved in the
D2-brane. Note that, under the assumption of time- and $z$-independence, 
closure of $F$ implies that $E$ must be constant but still allows  
$B$ to depend on $\sigma$. 

The supersymmetries of the IIA Minkowski vacuum 
preserved by a D2-brane configuration are those 
generated by (constant) Killing spinors 
$\e$ satisfying $\Gamma \e = \e$, where $\Gamma$ is the
matrix appearing in the `kappa-symmetry' transformation of the
D2-brane worldvolume spinors \cite{kappa}. 
For the configuration of interest here this condition reduces to
\be
y_i' \, \Gamma_{i}  \, \Gn \, 
\left( \Gamma_{TZ} \Gn + E \right) \, \e  
+ \left( B \, \Gamma_{T}\Gn - 
\sqrt{(1-E^2) |\vec{y}\,'|^2 + B^2} \right) \, \e =0 \,,
\label{susy}
\ee
where $\Gamma_T, \Gamma_Z$ and $\Gamma_i$ are (constant)
ten-dimensional Minkowski spacetime Dirac matrices, $\Gn=\Gamma_T
\Gamma_Z \Gamma_1 \ldots \Gamma_8$ is the
chirality matrix in ten dimensions, 
and  the prime denotes differentiation with respect to $\sigma$.

The supersymmetry preservation equation \eqn{susy} is satisfied for any arbitrary
curve provided that we set $|E|=1$, impose the two conditions 
\be
\Gamma_{TZ} \Gn \, \e = -\mbox{sgn} (E) \, \e \sac
\Gamma_{T}\Gn \, \e = \mbox{sgn} (B) \, \e 
\label{projections}
\ee
on the Killing spinors, and demand that $B(\sigma)$ is a
constant-sign, but otherwise completely arbitrary, function of
$\sigma$. We would like to emphasize that $|E|=1$ is 
{\it not} a critical electric field in the presence of a
nowhere-vanishing magnetic field $B(\sigma)$, 
which is the only case that we will consider in this paper. 

The two conditions \eqn{projections} on $\e$  correspond to string
charge along the $Z$-direction and to D0-brane charge, respectively,
and preserve 1/4 of the supersymmetry\footnote{The equation \eqn{susy}
admits other types of solutions for particular forms of the
cross-section. For example, if this is a straight line and $B$ 
is constant, then the fraction of preserved
supersymmetry is 1/2, corresponding to an infinite planar D2-brane
with a homogeneous  density of bound strings and D0-branes. In this
paper we will concentrate on the generic case of an arbitrary
cross-section.}. Note that the D2-brane projector does not appear; we
shall return to this issue later. It is straightforward to check that this
configuration satisfies the equations of motion derived from the 
D2-brane action, which (for unit surface tension) is
\be
S_{D2} = \int dt\,  dz\,  d\sigma \, \call_{D2} = 
- \int dt\,  dz\,  d\sigma \, \sqrt{-\det (g+F)} \,,
\label{action}
\ee
where 
\be
ds^2(g) = -dt^2 + dz^2 + |\vec{y}\,'|^2 d\sigma^2
\ee
is the induced metric on the D2-brane worldvolume. Note that the
area element is
\be
\mathrm{\sqrt{\det g_{spatial}}} = |\vec{y}\,'| \,.
\label{det}
\ee

\subsection{Mechanical Stability: Momentum versus Tension}

We have concluded that a D2-brane with an arbitrary shape 
and an arbitrary magnetic field $B(\sigma)$ preserves 1/4 of the
supersymmetries of the IIA vacuum, and therefore must be stable. In
`normal' circumstances this would be impossible because of the
D2-brane tension. In the present case this tension is exactly
balanced by the `centrifugal' force associated to the linear momentum 
density carried by the supertube, as we show below. The origin of this
momentum is not time-dependence (which would generically be
incompatible with supersymmetry) but the Poynting vector generated by
the crossed static electric and magnetic fields on the D2-brane.

The linear momentum density in the $i$-direction is easily
computed by momentarily allowing $Y^i$ to depend on time,
evaluating 
\be
\calp_i = \fc{\pa \call_{D2}}{\pa \dot{Y}^i} \,,
\ee
where the dot denotes differentiation with respect to time, and
finally setting $\dot{Y}^i=0$. The result is
\be
\calp_i = \fc{B E y_i'}{\sqrt{(1-E^2)|\vec{y}\,'|^2 + B^2}} \,.
\label{pxpy}
\ee
Similarly, the momentum conjugate to the electric field or `electric
displacement' is 
\be
\Pi (\sigma) = \fc{\pa \call_{D2}}{\pa E} = 
\fc{E |\vec{y}\,'|^2}{\sqrt{(1-E^2) |\vec{y}\,'|^2 + B^2}} \,.
\label{pi}
\ee

For supersymmetric configurations $|E|=1$ and the momenta \eqn{pxpy}
become
\be
\calp_i = \sgn(\Pi B) \, y_i' \,.
\label{pxpy-susy}
\ee
This combined with \eqn{pi} yields the condition
\be
|\vec{\calp}|^2 = |\Pi B| \,.
\label{pib-susy}
\ee
Note that the densities $\Pi(\sigma)$, $B(\sigma)$ and
$\calp_i(\sigma)$ depend on the parametrization of the curve
$\vec{y}(\sigma)$. However, the physical densities per unit D2-brane
area are reparametrization-invariant. Using \eqn{det} these are
\be
\mathrm{\Pi_{ph}(\sigma)} = \fc{\Pi(\sigma)}{|\vec{y}\,'|} \sac
\mathrm{B_{ph}(\sigma)} = \fc{B(\sigma)}{|\vec{y}\,'|} \sac
\calp^i_{\mathrm{ph}}(\sigma) = \fc{\calp^i(\sigma)}{|\vec{y}\,'|} \,.
\ee
Note that, by virtue of \eqn{pxpy-susy}, the momentum density per unit
area has constant unit magnitude, that is,  
$|\vec{\calp}_\mathrm{ph}| = 1$.

The stability of the supertube with an arbitrary shape may now be 
understood in `mechanical' terms as follows. The conserved string
charge and the D0-brane charge per unit length in the $Z$-direction 
carried by the supertube are (for an appropriate choice of units)
\be 
q_{F1} =  \int d\sigma \, \Pi \sac 
q_{D0} =  \int d\sigma \, B \,.
\label{charges}
\ee
The string and D0-brane charge densities $\Pi(\sigma)$ and $B(\sigma)$
generate a Poynting linear momentum density $\vec{\calp}$ 
at each point along the cross-section. For specified shape
and magnetic field $B$, supersymmetry automatically adjusts $\Pi$ 
in such a way that the momentum density per unit area 
is tangent to the curve and has constant magnitude. 
The fact that it is tangent means that,  in a
mechanical analogy, one can think of this momentum as originating from
a  continuous motion of the curve along itself, similarly to that of a
fluid along a tube of fixed shape.  The fact that it is constant in
magnitude means that the force acting on  each point of the curve must
be orthogonal to the momentum: this is precisely the force due to the
D2-brane tension. In the absence of momentum, this tension would make
the curve collapse, whereas here it provides precisely the required
centripetal force to direct the motion  of each point on the curve in
such a way that the curve is mapped  into itself under this
motion. This is how stability is achieved for an arbitrary shape.

One consequence of the precise balance of forces discussed above is
that the D2-brane behaves in a certain sense as a {\it tensionless}
brane. To see this, it is important to remember that the tension of a 
system is {\it not} the same as its energy per unit volume, and also that it
may be non-isotropic; the supertube is an example of this. 
The tension tensor is defined as minus the purely spatial part of the 
spacetime stress-energy tensor. The latter is computed as
\be
T^{MN}(x) = \left. \fc{2}{\sqrt{-\det G}} \, \fc{\d S_{D2}}{\d G_{MN}(x)} 
\right|_{G_{MN}=\eta_{MN}} \quad (M,N=0, \ldots, 9) 
\label{st-tensor}
\ee
by momentarily allowing for a general spacetime metric in the
D2-brane action and then setting it to its Minkowski value \eqn{st-metric}
after evaluating the variation above. The result is 
\be
T^{MN}(x) = \int d^3 \xi \, \calt^{MN}(X(\xi)) \, \delta^{(10)} (x - X(\xi)) \,,
\ee
where 
\be
\calt^{MN} = - \sqrt{-\det(g+F)} \, \left[ (g+F)^{-1} \right]^{(ab)}
\pa_a X^M \pa_b X^N 
\ee
and $X^M = \{T, Z, Y^i\}$. Note that $T^{MN}(x)$ is conserved, that
is, $\pa_M T^{MN} = 0$, by virtue of the D2-brane equations of motion 
\be
\pa_a \, \left( \sqrt{\det(g+F)} \, \left[ (g+F)^{-1} \right]^{ab}
\pa_b X^M \right) =0 \,.
\ee
Specializing to the supertube configurations of interest here we find
that the only non-zero components of $\calt^{MN}$ are
\be
\calt^{TT} = |\Pi| + |B| \sac \calt^{ZZ} = - |\Pi| \sac 
\calt^{Ti} = \mbox{sgn}(\Pi B) \, y_i' \,.
\ee
This result illustrates a number of points. First, the off-diagonal
components $\calt^{Ti}$ are precisely the linear momentum density
\eqn{pxpy-susy} carried by the tube, as expected. Second, the fact that  
$\calt^{ij}$ vanishes identically means that there is no tension 
along the cross-section; this provides a more formal explanation of why an
arbitrary shape is stable. Third, the tube tension $-\calt^{ZZ}=|\Pi|$ 
in the $Z$-direction is only due to the string density. 
Hence the D2-brane does not contribute to the tension in any
direction: it has effectively become tensionless. 
(The fact that the D0-branes do not contribute to the tension either 
should be expected: they behave like dust.) Finally, the net energy 
of the supertube per unit length in the $Z$-direction
\be
\calh = \int d\sigma \, \calt^{TT} = |q_{F1}| + |q_{D0}| 
\label{energy}
\ee
saturates the lower bound which we will derive  from the
supersymmetry algebra below. In particular, it receives contributions 
from the strings and the D0-branes, but not from the D2-brane.
The reason for this is that the supertube is a true bound state
in which the strictly negative energy which binds the strings and
D0-branes to the D2-brane is exactly cancelled by the strictly 
positive energies associated to the mass of the D2-brane 
and to the presence of a linear momentum density.

\subsection{Central Charges}

Supersymmetric configurations in a given theory are those which
minimize the energy for fixed values of the central charges in the 
supersymmetry algebra of the theory. Both this energy and the precise set
of preserved supersymmetries are completely determined once
the central charges are specified. For the type IIA theory, 
the anti-commutator of two supercharges in the presence of the central
charges $\calz$ of interest here is
\be
\{ Q, Q \} = \Gamma^T \Gamma^M P_M + 
\fc{1}{2} \Gamma^T \Gamma^{MN} \calz^{D2}_{MN} + 
\Gamma^T \Gamma^{M} \Gn \calz^{F1}_{M} + \Gamma^T \Gn \calz^{D0}  \,.
\label{algebra}
\ee
In the previous sections we have understood the supersymmetries 
preserved by the supertube and its mechanical stability from a local 
analysis. One purpose of this section is to show that this also
follows from consideration of the central charges in the algebra above
carried by the supertube. We shall also prove the existence of upper 
bounds on the total angular momentum and on the linear momentum density. 

By definition, any supertube carries non-zero string and D0-brane
charges. In addition, it may carry a total (per unit length in the
$Z$-direction) linear momentum $P_i$ and/or angular momentum 2-form 
$L_{ij} = - L_{ji}$ in $\bbe{8}$.
These are obtained by integrating the corresponding densities along
the cross-section. For $N$ overlapping D2-branes the linear momentum 
density \eqn{pxpy-susy} for supersymmetric configurations becomes
\be
\calp_i = N \, y_i ' \,,
\label{P-N-y}
\ee
(where we have chosen $\mbox{sgn}(\Pi B)=+1$ for concreteness)
and hence the total linear momentum is
\be
P_i = N \, \int d\sigma \, y_i' = N \, \int dy_i \,.
\label{linear}
\ee
Note that the momentum density per unit area has magnitude $N$, that is,
\be
|\mathrm{\vec{\calp}_{ph}}| = N \,.
\label{magnitude}
\ee
Similarly, the angular momentum density is
\be
\call_{ij} \equiv Y_i \, \calp_j -  Y_j \, \calp_i \,,
\ee
so in the supersymmetric case the total angular momentum takes the form
\be
L_{ij} = N \, \int d\sigma \, \left( y_i \, y_j' - y_j \, y_i' \right)
= N \, \int \left( y_i \, dy_j - y_j \, dy_i \right) \,.
\label{angular}
\ee

Consider now a supertube with a closed cross-section. In this case the
angular momentum in a given $ij$-plane is precisely the number of
D2-branes times twice the area of the region $\cala$ enclosed by the
projection $\calc$ of the cross-section onto that plane, since by
Stoke's theorem we have
\be
L_{ij} = N \, \oint_{\calc} \left( y_i \, dy_j - y_j \, dy_i \right) =
2N \, \int_{\cala} \, dy_i \wedge dy_j \,.
\ee
Consequently, a non-zero angular momentum prevents a closed
cross-section from collapse, since the area enclosed by a collapsed
cross-section would vanish. A closed supertube is therefore stabilized
by the angular momentum. It may appear surprising that this can be
done {\it supersymmetrically} in the Minkowski vacuum, since angular
momentum  is not a central charge in Minkowski supersymmetry\footnote{
This should be contrasted with the case of $AdS$ supersymmetry,  for
which the angular momentum {\it is} one of the central charges.
Presumably, this is related to the supersymmetry of the giant
gravitons \cite{giant}.}. There is no contradiction, however, because
the set of preserved supersymmetries and the energy of a closed
supertube are independent of the angular momentum. In fact, they are
precisely the same as those of a supersymmetric collection of strings
and D0-branes (see \eqn{projections} and \eqn{energy}). This follows
from the supersymmetry algebra because both systems carry the
same central charges: for a closed supertube, the total linear 
momentum \eqn{linear} vanishes because in this case $y_i(\sigma)$ 
are periodic, and the net D2-brane charge vanishes because one of the 
D2-brane directions is compact.  

The case of a supertube with an open cross-section which extends
asymptotically to infinity\footnote{This must indeed be the case for a
non-closed curve, since by charge conservation a D2-brane cannot 
have a boundary unless it ends on another appropriate brane.} 
is more subtle, because in this case there is both a total linear 
momentum and a net D2-brane charge. Nevertheless, the supersymmetry 
algebra still implies the same bound on the energy and the same
supersymmetry conditions because of a `cancellation' between the 
momentum and the D2-brane charge in the supersymmetry algebra. 
Indeed, preserved supersymmetries correspond to eigen-spinors $\e$ 
of the matrix in \eqn{algebra} with zero eigen-value. 
Specifying to a supertube with cross-section extending asymptotically along a
direction denoted by $\|$ equation \eqn{algebra} becomes
\be
\{ Q, Q \} = P^0 + N \Gamma_T \Gamma_\| 
\left( 1 - \Gamma_Z \right) + 
\Gamma_{TZ} \Gn \calz_{F1} + \Gamma_{T} \Gn \calz_{D0} \,,
\ee
where we have made use of the fact that $P_\|= N$ (see
\eqn{magnitude}) and that $\calz^{D2}_{z\|}=N$. Now it is clear that if 
we impose the conditions \eqn{projections} on $\e$ then the 
term in brackets originating from the linear momentum and the 
D2-brane charge automatically vanishes, and that in order for 
$\e$ to have zero eigen-value we must have 
\be
P^0 = |\calz_{F1}| + |\calz_{D0}| \,.
\ee
This is essentially the integrated (in the $Z$-direction) version of equation
\eqn{energy}, and the usual arguments show that the right-hand side 
is a lower bound on the energy of any configuration with the same
charges. 

We now turn to the upper bound on the angular momentum. For
simplicity, here we restrict ourselves to closed supertubes. 
We choose the parametrization such that $\vec{y}(\sigma)
=\vec{y}(\sigma+1)$, so that the angular momentum 2-form per period is
\be
L_{ij} = N \, \int_0^1 d \sigma \, 
\left(y_i \, y_j' - y_j \, y_i' \right) \,.
\ee
The total angular momentum $J$ is defined as 
\be
J \equiv \sqrt{\fc{1}{2} L_{ij} L^{ij}} \,,
\label{J}
\ee
and satisfies the bound\footnote{This bound was derived in \cite{EMT01} for
a supertube with a circular cross-section.}
\be
J \leq N^{-1} \, |q_{F1} \, q_{D0}| \,.
\label{Jbound}
\ee
To see this, we assume (without loss of generality) that the $\bbe{8}$
axes are oriented such that the angular momentum 2-form $L$ is
skew-diagonal with skew-eigenvalues $\ell_\a = L_{2\a-1,2\a}$, 
$\a=1, \ldots, 4$, and that the cross-section is parametrized in 
pairs of polar coordinates as  
\bea
y^{2\a -1} &=& R_\a (\sigma) \, \sin \left( 2 \pi \sigma \right) \,, \nn 
y^{2\a} &=& R_\a (\sigma) \, \cos \left( 2 \pi \sigma \right) \,,
\label{polar}
\eea
where $R_\a$ are four position-dependent radii.
Integrating over a period we have the following chain of (in)equalities:
\bea
J &=& \left( \sum_{\a=1}^4 \ell_a^2 \right)^{1/2} \nn
&=& N \, \left( \sum_{\a=1}^4 \left[ 
\int_0^1 d\sigma \, R_\a^2(\sigma) \right]^2 \right)^{1/2} \nn
&\leq& N \, \sum_{\a=1}^4 \int_0^1 d\sigma \, R_\a^2(\sigma) \nn
&\leq& N \, \sum_{\a=1}^4 \int_0^1 d\sigma \, 
\left(R_\a^2(\sigma) + R'^2_\a(\sigma) \right) \nn
&=& N \, \sum_{\a=1}^4 \int_0^1 d\sigma \, 
\left( |y'_{2\a -1}(\sigma)|^2 + |y'_{2\a}(\sigma)|^2 \right) \nn
&=& N \, \int_0^1 d\sigma \, |\vec{y}\, '(\sigma)|^2  \nn
&=& N^{-1} \, \int_0^1 d\sigma \, |\Pi(\sigma) \, B(\sigma)|  \nn
&\leq&  N^{-1}\, \left( \int_0^1 d\sigma \, |\Pi(\sigma)| \right) 
\left( \int_0^1 d\sigma \, |B(\sigma)| \right) \nn
&=&  N^{-1} \, |q_{F1} \, q_{D0}| \,,
\label{inequalities}
\eea
where we have made use of \eqn{J}, \eqn{angular}, \eqn{polar},
\eqn{P-N-y}, \eqn{pib-susy} and \eqn{charges}. 
As may be seen by demanding the saturation of all the inequalities
in \eqn{inequalities}, equality in \eqn{Jbound} is achieved if and
only if: ({\it i}) all but one of the skew-eigenvalues of $L$ vanish, so the
supertube cross-section is a curve in $\bbe{2}$, 
({\it ii}) this curve is a circle, so the supertube is a perfect
cylinder, and ({\it iii}) $\Pi$ and $B$ are constant, so the supertube
carries homogeneous string and D0-brane charge densities.

We finally turn to the linear momentum density. For a general
1/4-supersymmetric type IIA configuration, the magnitude of
$\vec{\calp}_\mathrm{ph}$ is not actually given by \eqn{pib-susy}, 
but instead is only bounded from above by the right-hand-side of this
equation; this observation is important for comparison with the 
supergravity description. The reason is that the magnitude of the 
string and D0-brane charge densities at any point along the 
cross-section of a supertube can be increased without increasing 
the momentum density and while preserving supersymmetry. This is
because, as discussed exhaustively in \cite{EMT01} for a circular
supertube and as we shall see in Section \ref{sugra} for a general one,
there is no force between a supertube and 
strings (aligned along the $Z$-direction) or D0-branes (with charges
of the same sign as those on the tube) placed at rest at arbitrary 
distances from each other; the combined system still preserves 1/4
supersymmetry. In particular, these extra strings and D0-branes can 
be superposed with the tube without being bound to it, hence 
increasing $|\Pi B|$ but leaving $|\vec{\calp}|$ intact. 
Thus for a general combined system we conclude that 
\be
|\calp|^2 \leq |\Pi B| \,.
\label{bound}
\ee


\sect{Supermembrane Analysis}
\label{membrane}


We have just seen that the cross-section of a D2-brane supertube may have an
arbitrary shape.  We shall now show that this result can be understood from an
M-theory perspective via the light-front-gauge supermembrane. This analysis
provides a link between the DBI approach and the Matrix model approach,
used in \cite{BL01} to recover the supertube with circular
cross-section and then generalized in \cite{BK01} to a supertube with 
elliptical cross-section. The reason that the two
approaches are related is that the D2-brane DBI action in a IIA vacuum
background is equivalent to the action for the supermembrane  action 
in a background spacetime of the form $\bbe{(1,9)}\times S^1$; 
the former is obtained from the latter by dualization of the scalar 
field that gives the  position of the membrane on the $S^1$ factor. 
But the supermembrane action in  light-cone gauge is a
supersymmetric gauge quantum mechanics (SGQM) model with the group $SDiff$
of area-preserving diffeomorphisms of the membrane as its gauge group, 
and this becomes a Matrix model when $SDiff$ is approximated by 
$SU(N)$ \cite{dWHN}. Here we shall work directly with the $SDiff$
model, and take the decompactification limit. 

In choosing the light-front gauge, one first chooses the metric
\be
ds^2_{\it 11} = dY^+dY_- + \sum_{I=1}^9 dY^IdY^I\, .
\ee
The physical worldvolume fields $Y^I$ specify the position of the membrane
in a 9-dimensional space $\bbe{9}$. The bosonic light-front gauge
Hamiltonian density is
\be
{\cal H} = {1\over2}\left[ \sum_I P_I^2 + \sum_{I < J}
\{Y^I,Y^J\}^2\right]\, ,
\ee
where $P_I$ is the momentum space variable conjugate to $Y^I$. The bracket
$\{f,g\}$ of any two functions on the membrane is the Lie bracket
\be
\{f,g\} \equiv \varepsilon^{ab}\partial_a f \partial_b g\, ,
\ee
where $\sigma^a$ ($a=1,2$) are the (arbitrary) membrane coordinates.
Although the variables $(Y^I,P_I)$ span an 18-dimensional space, the
physical bosonic phase space is only 16-dimensional because of the
Gauss law constraint 
\be
\sum_I \{Y^I,P_I\} =0 \, ,
\ee
and the $SDiff$ gauge transformation that the constraint function generates.

We begin by separating the coordinates as $Y^I = ( Z , Y^i)$, 
where $Y^i$ are Cartesian coordinates on $\bbe{8}$, and their
corresponding conjugate momenta  as $P_I = (P_Z , P_i )$. The bosonic 
Hamiltonian density is now
\be
\label{hamden}
{\cal H} = {1\over 2} \left( \sum_i P_i^2 + P_Z^2 + 
\sum_{i < j} \{ Y^i,Y^j \}^2 + \sum_i \{Y^i,Z\}^2 \right) \, ,
\ee
and the Gauss-law constraint is
\be
\label{constraint}
\{Y^i,P_i\}  + \{Z,P_Z\} =0 \,.
\ee
We will be interested in solutions that preserve 8 of the 16 
supersymmetries of the $D$=11 supermembrane theory, 
and hence 1/4 of the supersymmetry of the M-theory vacuum.

We expect supersymmetric solutions of the membrane equations of motion to
minimize the energy for fixed conserved charges, and we shall use this
method to find them. Following \cite{BL01} we use (\ref{constraint})
to rewrite the Hamiltonian density as
\bea
\label{hamrewrite}
{\cal H} &=& {1\over2} \sum_i \left( P_i - \{Z,Y^i\}\right)^2 
+ \fc{1}{2} P_Z^2 + \fc{1}{2} \sum_{i < j} \{ Y^i,Y^j \}^2 \nn
&& +\ \varepsilon^{ab}\partial_a \, \left[
Z \, \left( P_i \partial_b Y^i + P_Z \partial_b Z \right) \right] \,.
\eea
For fixed boundary conditions at any boundary of the membrane, the energy
density is therefore minimized locally by configurations satisfying both
\be\label{BPS1}
\{Y^i,Y^j\} = 0 
\ee
and
\be\label{BPS2}
P_i = \{Z,Y^i\} \sac P_Z=0 \,.
\ee
For configurations with the tubular topology of the previous section
the Hamiltonian is 
\be
H= {1\over2} \left[\int \{Z^2,Y^i\} \, dY^i \right] \,,
\ee
where the integral is over the cross-sectional curve in $\bbe{8}$ and
the square brackets indicate an evaluation `at the ends' of the
$Z$-direction. 

To verify that configurations satisfying (\ref{BPS1}) and (\ref{BPS2})
indeed preserve the expected fraction of supersymmetry, we make use of
the fact that the supersymmetry transformation of the 16-component 
worldvolume $SO(9)$ spinor field $S$ is
\be
\delta S = \left[P_I\gamma^I + {1\over2} \{Y^I,Y^J\}\gamma_{IJ}\right]
\alpha + \beta \,,
\ee
where $\alpha$ and $\beta$ are two 16-component constant $SO(9)$ spinor
parameters, and $\gamma^I$ are the $16\times 16$ $SO(9)$ Dirac matrices.
Clearly, all of the `$\beta$-supersymmetries' are broken, and the 16
components of $S$ are the corresponding Nambu-Goldstone fermions. Setting
$\beta$ to zero and making use of the relations (\ref{BPS1}) and 
(\ref{BPS2}), the supersymmetry transformation becomes
\be
\delta S = P_i \gamma_i \, \left(1 -\gamma_Z\right)\alpha \,,
\ee
where the three Dirac matrices $\gamma^i$ form a reducible 16-dimensional
representation of the Clifford algebra for $SO(8)$. 
This vanishes for parameters $\alpha$ satisfying
\be
\gamma_Z \alpha = \alpha \,,
\ee
which reduces the number of non-zero supersymmetry parameters to 8.

Let us now study some explicit solutions of the equations 
\eqn{BPS1} and \eqn{BPS2}. To make contact with \cite{BL01} 
we first concentrate on a circular cross-section of 
radius $R$ (which we initially allow to depend on the position along 
the axis of the tube) in the, say, 12-plane. Hence we choose membrane 
coordinates $(z,\vp)$ and set 
\be 
Y^1 = R(z) \cos \vp  \sac  Y^2 = R(z) \sin\vp  \sac  Z=z \,.
\ee
It now follows from \eqn{BPS1} that $R$ must be constant, in which
case $Y^1$, $Y^2$ and $Z$ satisfy 
\be
\{Z,Y^1\} = -Y^2 \sac \{Z, Y^2\} = Y^1  \sac  \{Y^1, Y^2\} = 0 \,,
\ee
which implies that the angular momentum in the $Z$-direction is
\be
L_Z= Y^1 P_2 - Y^2 P_1 = R^2 \,.
\ee
If we now replace the Poisson bracket of functions on the membrane by $-i$
times the commutator of $N\times N$  
Hermitian matrices, then we recover the Matrix theory 
description of the supertube given in \cite{BL01}. It is manifest 
from our derivation of this result that the large $N$ limit yields a 
tubular membrane with, in this case, a circular cross-section. 
It should be noted, however, that the
identification of this 1/4-supersymmetric tubular membrane 
with the D2-brane supertube is not immediate because the latter lifts to
a time-dependent M2-brane configuration in standard Minkowski 
coordinates \cite{MT01}. Moreover, there exist 1/4-supersymmetric
tubular solutions of the DBI equations for {\it non-constant} 
$R(z)$ \cite{MT01}, whereas supersymmetry forces constant $R$ in the
above Matrix model approach. Thus, the equations \eqn{BPS1} and \eqn{BPS2} do not
capture all possibilities for supersymmetric tubular D2-branes.
However, the method does capture the possibility of an arbitrary
cross-section. Indeed, choose membrane coordinates $(z,\sigma)$ and set 
\be
Y^i = y^i(\sigma) \sac Z=z \,.
\label{1}
\ee
An argument analogous to that which led to \eqn{pxpy-susy} shows
that we then have
\be
P_i = y_i' \,.
\label{2}
\ee
It is now immediate to verify that \eqn{1} and \eqn{2} solve the 
supersymmetry equations \eqn{BPS1} and \eqn{BPS2}.


\sect{Supergravity Solution}
\label{sugra}

The supergravity solution sourced by a supertube with arbitrary 
cross-section takes the form
\bea
ds^2_{\it 10} &=& - U^{-1} V^{-1/2} \, ( dT - A)^2 +
U^{-1} V^{1/2} \, dZ^2 + V^{1/2} \, d\vec{Y} \cdot d\vec{Y} \,, \nn
B_{\it 2} &=& - U^{-1} \, (dT - A) \wedge dZ + dT \wedge dZ\,, \nn
C_{\it 1} &=& - V^{-1} \, (dT - A) + dT \,, \label{solution} \\
C_{\it 3} &=& - U^{-1} dT\wedge dZ \wedge A \,, \nn
e^\phi &=& U^{-1/2} V^{3/4} \,, \nonumber
\eea
where, as above, $\vec{Y}$ are Cartesian coordinates
on $\bbe{8}$. $U(\vec{Y})$ and $V(\vec{Y})$ are harmonic functions on
$\bbe{8}$, and $A(\vec{Y})$ is a 1-form on $\bbe{8}$ which 
must satisfy Maxwell's equation $d *_8 dA =0$. 
$B_{\it 2}$ and $C_{\it p}$ are the Neveu-Schwarz and
Ramond-Ramond potentials, respectively, with gauge-invariant
field-strengths
\be
H_{\it 3}=dB_{\it 2} \sac F_{\it 2}=dC_{\it 1} \sac 
G_{\it 4} = dC_{\it 3} - dB_{\it 2} \w C_{\it 1} \,.
\ee
Note for future reference that for the solution above 
\be
G_{\it 4} = U^{-1} V^{-1} (dT -A) \w dZ \w dA \,.
\label{g4}
\ee

The  solution \eqn{solution} was actually presented in \cite{EMT01},
but there only the choice of $U$, $V$ and $A$ that describes a supertube
with circular cross-section was found. The generalization to an
$N$-D2-brane tube with an {\it arbitrary} cross-section in 
$\bbe{8}$ specified by $\vec{Y} = \vec{y}(\sigma)$ and carrying 
string and D0-brane charge densities $\Pi(\sigma)$ and $B(\sigma)$ is
\bea
U(\vec{Y}) &=& 1 + \fc{1}{6 \Omega_{\it 7}} \, \int d \sigma \, 
\fc{|\Pi(\sigma)|}{| \vec{Y} - \vec{y}(\sigma)|^6}  \,, \label{U} \\
V(\vec{Y}) &=& 1 + \fc{1}{6 \Omega_{\it 7}} \, \int d \sigma \, 
\fc{|B(\sigma)|}{| \vec{Y} - \vec{y}(\sigma)|^6}  \,, \label{V} \\
A(\vec{Y}) &=& \fc{N}{6 \Omega_{\it 7}} \, \int d\sigma \, 
\fc{y_i'(\sigma)}{| \vec{Y} - \vec{y}(\sigma)|^6} \, dY^i
\,, \label{A} 
\eea
where $\Omega_{\it q}$ is the volume of a unit $q$-sphere.
The solution \eqn{solution} with these choices correctly reproduces
all the features of the supertube. First, it preserves the same
supersymmetries, since for any choice of $U$, $V$ and $A$, the
solution \eqn{solution} is invariant under eight supersymmetries
generated by Killing spinors of the form (in the obvious orthonormal
frame for the metric)
\be
\e= U^{-1/4} V^{-1/8} \e_0 \,,
\ee
where $\e_0$ is a constant spinor which must satisfy precisely the
constraints \eqn{projections}\footnote{With
$\mbox{sgn}(E)=\mbox{sgn}(B)=1$; these signs are reversed by taking 
$T \rightarrow -T$ and/or $Z \rightarrow -Z$.}
\cite{EMT01}. 

Second, it carries all the appropriate charges. To see this, we need
to distinguish between closed and open cross-sections. Suppose first
that the cross-section is closed. In this case we shall assume that it
is contained in a compact region of $\bbe{8}$ of finite size 
and that the charges \eqn{charges} and momenta \eqn{linear} and 
\eqn{angular} are finite. Under these circumstances we have the
following asymptotic behaviour for large $|\vec{Y}|$:
\bea
U(\vec{Y}) &\sim& 1 + \fc{|q_{F1}|}{6 \Omega_{\it 7} |\vec{Y}|^6} +
\cdots \,, \nn
V(\vec{Y}) &\sim& 1 + \fc{|q_{D0}|}{6 \Omega_{\it 7} |\vec{Y}|^6} +
\cdots \,, \nn
A(\vec{Y}) &\sim& \fc{L_{ij} Y^j}{2 \Omega_{\it 7} |\vec{Y}|^8} 
\, dY^i + \cdots \,,
\eea
where the dots stand for terms sub-leading in the limit under consideration
and $q_{F1}$, $q_{D0}$ and $L_{ij}$ are defined as
in \eqn{charges} and \eqn{angular}. This shows that the metric is
asymptotically flat for large $|\vec{Y}|$. We see
from the contributions of $U$ and $V$ to the field-strengths
$H_{\it 3}$ and $F_{\it 2}$ that the
solution carries string charge $q_{F1}$ and D0-brane charge per unit
length in the $Z$-direction $q_{D0}$. Evaluation of the appropriate
ADM integrals shows that it also carries an energy per unit length in
the $Z$-direction
\be
\calh = |q_{F1}| + |q_{D0}| \,,
\ee
as well as angular momentum $L_{ij}$ in the $ij$-plane. Note that
there is no linear momentum, as expected for a closed curve.
The field-strength \eqn{g4} sourced by the D2-brane takes the asymptotic form 
\be
G_{\it 4} \sim d \left[ \fc{L_{ij} Y^j}{2 \Omega_{\it 7} 
|\vec{Y}|^8} \, dT \w dZ \w dY^i \right] + \cdots \,.
\ee
Since the integral of $*G_{\it 4}$ over any 6-sphere at infinity
vanishes, the solution carries no net D2-brane charge, as expected for a
closed supertube. However, as explained in \cite{EMT01}, the fact that 
$G_{\it 4}$ does not vanish implies that the D2-brane dipole (and higher)
moments are non-zero.

Consider now an open cross-section. For simplicity we assume that 
outside some compact region the curve becomes a straight line along
the, say, $Y^1$-direction, and that the densities per unit area 
$\mathrm{\Pi_{ph}(\sigma)}$ and $\mathrm{B_{ph}(\sigma)}$
become constants $\Pi_0$ and $B_0$. Under these circumstances we have
the following asymptotic expansions 
\bea
U &\sim& 1 + \fc{|\Pi_0|}{5 \Omega_6 |\vec{Y}_\perp|^5} + \cdots \,, \nn
V &\sim& 1 + \fc{|B_0|}{5 \Omega_6 |\vec{Y}_\perp|^5} + \cdots \,, \nn
A &\sim& \fc{N}{5 \Omega_6 |\vec{Y}_\perp|^5} \, dY^1 + \cdots \,,
\eea
for large $|\vec{Y}_\perp|$, where  
\be
\vec{Y}_\perp=(Y^2, \ldots, Y^8)
\label{perp}
\ee
is the position in the transverse directions. These expansions show
that, at large distances, the solution describes $N$ infinite D2-branes 
that extend asymptotically along the $ZY^1$-plane, with strings (along the
$Z$-direction) and D0-branes bound to them. Indeed, the solution is
asymptotically flat at tranverse infinity, that is, 
for large $|\vec{Y}_\perp|$. 
From the asymptotic forms of $H_{\it 3}$ and $F_{\it 2}$ 
we see that it carries string and D0-brane charge densities 
$\Pi_0$ and $B_0$, respectively. The energy density 
is $|\Pi_0| + |B_0|$. We also see from the 
asymptotic behaviour of the `$dT dY^1$-term' in the
metric that, unlike in the closed case, there are now $N$ units of 
net linear momentum in the 1-direction per unit area.
Similarly, now there are $N$ units of net D2-brane charge, 
since asymptotically we have
\be
G_{\it 4} \sim d \left[ \fc{N}{5 \Omega_6 |\vec{Y}_\perp|^5} \,  
dT \w dZ \w dY^1 \right] + \cdots \,. 
\ee
This is precisely the long-distance field-strength associated to $N$
D2-branes oriented along the $ZY^1$-plane.

The third feature which allows the solution with the
choices \eqn{U}-\eqn{A} to be identified with the supertube is that
(by construction) it is singular at and only at the
location of the tube $\vec{Y} = \vec{y}(\sigma)$. In fact, the
behaviour of the solution in the region close to the singularity 
will allow us to reproduce the bound \eqn{bound}.
Let $\vec{Y}_0=\vec{y}(\sigma_0)$ be a point on the
tube, $\vec{v} =\vec{y}\,'(\sigma_0)$ the tangent vector and 
$\Pi_0=\Pi(\sigma_0)$ and $B_0=B(\sigma_0)$
the string and D0-brane densities at that point.
By performing an appropriate translation and rotation if necessary we 
assume that $\vec{Y_0}=0$ and that $\vec{v}$ lies along the 
1-direction. It is now straightforward to see that in the limit 
$\vec{Y} \rightarrow \vec{Y}_0$ we have
\bea
U &\sim& 1 + \fc{|\Pi_0|}
{5 \Omega_6 |\vec{v}| |\vec{Y}_\perp|^5} + \cdots \,, \nn
V &\sim& 1 + \fc{|B_0|}
{5 \Omega_6 |\vec{v}| |\vec{Y}_\perp|^5} + \cdots \,, \nn
A &\sim& \fc{|\vec{\calp}|}
{5 \Omega_6 |\vec{v}| |\vec{Y}_\perp|^5} \, dY^1 + \cdots \,,
\label{expansion}
\eea
where $|\vec{\calp}|=N |\vec{v}|$, with $\vec{Y}_\perp$ as 
in \eqn{perp} and where the dots stand for $\vec{Y}_\perp$-dependent 
subleading terms. We see that in this limit
the tube behaves as an infinite planar D2/F1/D0-bound state extending
along the $ZY^1$-plane which carries a momentum density $\vec{\calp}$
along the 1-direction. The bound on this density arises now as
follows. Consider the vector field 
$\ell \equiv \pa / \pa Y^1$, which becomes tangent to the curve as 
$\vec{Y} \rightarrow \vec{Y}_0$. Its norm squared is
\be
|\ell|^2 = U^{-1} V^{-1/2} \, \left( UV - A_1^2 \right) \,.
\label{norm}
\ee
This is always positive sufficiently far away from the tube, and if
the bound \eqn{bound} is satisfied it remains spacelike
everywhere. However, if the bound is violated then $\ell$
becomes timelike sufficiently close to the point $\vec{Y}_0$. To see 
this we note that, as $\vec{Y} \rightarrow \vec{Y}_0$, \eqn{norm} becomes
\be
|\ell|^2 = \left( 5 \Omega_6 |\vec{Y}_\perp|^5 \right)^{-1/2} \, 
|\Pi_0|^{-1} \, |B_0|^{-1/2}\, |\vec{v}|^{-2} \, \Big[
\left( |\Pi_0 B_0| - |\calp|^2 \right) + \cdots 
\Big] \,,
\ee
where the dots stand for non-negative terms which vanish in this limit. Thus
$|\ell|^2$ becomes negative sufficiently close to $\vec{Y}_0$ if and
only if $|\calp|^2 > |\Pi_0 B_0|$. If this happens at every point
along the cross-section then curves almost tangent and sufficiently
close to the supertube are timelike. For a closed cross-section this
leads to a global violation of causality since these become closed
timelike curves. This case was analyzed in detail in \cite{EMT01} for
a circular cross-section, where it was shown that regions with
timelike $\ell$ also lead to another pathology: the appearance of
ghost degrees of freedom on appropriate brane probes. The reason is
that the coefficient of the kinetic energy of certain fields on the probe
is proportional to $|\ell|^2$. Since this is a local instability, it
will still occur even if the bound \eqn{bound} is violated only locally
and no closed timelike curves are present. Hence, it is the
requirement of stability of brane probes that leads to the bound 
\eqn{bound} in the supergravity description of supertubes with
arbitrary cross-sections.


\section{Absence of Tachyons}
\label{string}


In this section we compute the vacuum energy of the open strings
stretched between tangent planes of the supertube and show that no
tachyons are present. We follow the light-cone gauge 
boundary state formalism of \cite{GG96} closely, although in that
paper the formalism was developed explicitly only for type IIB
D-branes. Here we provide the slight modification necessary to treat
the type IIA D-branes.


\subsection{Generalities}


The boundary states constructed in \cite{GG96} in the light-cone-gauge
formalism describe `$(p+1)$-instantons' with Euclidean
worldvolume. Since each such $(p+1)$-instanton is related to an
ordinary D$p$-brane with Lorentzian worldvolume by a double Wick 
rotation we shall just use the term D$p$-brane in the following.

In the light-cone description the spacetime coordinates are 
divided into the light-cone coordinates $X^+ ,X^-$ and
the transverse coordinates $X^I$ ($I=1, \ldots , 8$). We shall further
separate the latter into coordinates $X^i$ ($i=1, \ldots , p+1$) 
parallel to the D$p$-brane and coordinates $X^n$ ($n=p+2, \ldots, 8$) 
transverse to it.

The boundary state $\ket{B}$ corresponding to an infinite planar 
D$p$-brane or anti-D$p$-brane is completely determined by the 
orientation of the brane in spacetime and by the worldvolume BI 2-form 
field-strength $F_{ij}$. This information is encoded in an 
$O(8)$ `rotation' constructed as follows. 

Consider first a D$p$-brane extended along the directions 
$1,\ldots,p$. Define the $O(8)$ matrix 
\be
M_{IJ} = \pmatrix { M_{ij} &  \cr
                             &   \bbi{7-p}  \cr} \,,
\label{mIJ}
\ee
where $\bbi{q}$ is the $(q \times q)$-dimensional identity  matrix and
\be
M_{ij} = - \left[(1-F)(1+F)^{-1} \right]_{ij} \,.
\ee
Note that in the type IIB theory $p$ is odd and hence $\det M =1$; this is the
case considered explicitly in \cite{GG96}. We shall instead
concentrate on the type IIA theory, for which  $p$ is even and hence 
$\det M=-1$. In this case it is convenient to write $M$ as
\be
M = \cald \cdot \tilde{M} \,, 
\ee
where 
\be
\cald \equiv \pmatrix {  - \bbi{p+1} &  \cr
                             &   \bbi{7-p}  \cr} \sac
\tilde{M} \equiv \cald \cdot M \,.
\ee
Note that $\cald^2 =1$ and $\det \tilde{M} = 1$.

The rotation $M_{IJ}$ gives rise to two different elements
in the spinor representation differing by a sign. In particular,
if we write 
\be
\tilde{M}_{IJ} = \exp \left(\fc{1}{2}\Omega_{KL} 
\Sigma^{KL}\right)_{IJ} \,,
\ee
where $\Sigma^{KL}$ are the generators of $SO(8)$ in the vector 
representation and $\Omega_{KL}=-\Omega_{LK}$ are constants, then 
these two elements are
\be
M_{ab} = \pm \left( \cald \cdot \tilde{M} \right)_{ab} \,,
\label{Maa}
\ee
where\footnote{Although in the type IIA theory the left-moving and the
right-moving spinors have opposite chiralities, we shall not
explicitly distinguish between chiral and anti-chiral spinor indices, 
that is, between dotted and undotted indices.} 
$a,b =1 , \ldots , 8$, $\gamma^{I}$ are $SO(8)$ Dirac gamma-matrices and
\bea
\cald_{ab} &=&  \left( 
\gamma^{p+2} \gamma^{p+3} \cdots \gamma^8 \right)_{ab} \,, \nn
\tilde{M}_{ab} &=& 
\exp \left( \fc{1}{4} \Omega_{IJ} \g^{IJ} \right)_{ab} \,.
\eea
The two possible choices of sign in \eqn{Maa} correspond to the 
possibilities, for fixed orientation and BI field-strength, of a
D-brane or an anti-D-brane. 

Consider now a D$p$-brane with an arbitrary orientation obtained 
from the previous one by an $SO(8)$ rotation $m(\phi)$, specified 
by some angles collectively denoted by $\phi$. Then the corresponding
matrix $M(\phi)$ is 
\be
M(\phi) = m^{-1}(\phi) \cdot M \cdot m(\phi) \,.
\label{conjugation}
\ee
In particular, when $F=0$, a rotation of angle $\phi=\pi$ on a 
2-plane with one direction parallel to the brane and another direction
orthogonal to it leaves $M_{IJ}$ invariant but reverses the sign of
$M_{ab}$, and hence it transforms a D-brane into an anti-D-brane and
viceversa. This is the generalization to branes of any dimension
of the familiar fact that reversing the orientation of a string
tranforms it into an anti-string.

The boundary state is completely specified once $M$ is known, and 
takes the form
\be
\ket{B} = \d^{(p+1)}_\perp \, R(\tilde{M}) \exp \sum_{n>0} \left({1\over n}
\cald_{IJ} \alpha^I_{-n}\tilde{\alpha}^J_{-n}  -
i \cald_{ab} S^a_{-n}\tilde{S}^{b}_{-n}\right) \ket{B_0} \,,
\label{state}
\ee
where the delta function above has support on the worldvolume of the
brane (see the Appendix for details), and the zero-mode factor is
\be
\ket{B_0} = C \left(   M_{IJ} \ket{I}\ket{J}  +
i M_{ab} \ket{a}\ket{b}  \right) \,.
\ee
The normalization constant $C$ is given in terms of the BI
field-strength on the D-brane as
\be
C(F) = \sqrt{\det(1+F)} \,.
\ee
Finally, the operator $R(\tilde{M})$ is the representation of the 
$SO(8)$-rotation $\tilde{M}$ on non-zero modes:
\be
R(\tilde{M}) = \exp \sum_{n>0} \left({1\over n}  T^{(\a)}_{IJ}
\alpha^I_{-n}{\a}^J_{n} +   
T^{(S)}_{ab} S^a_{-n}{S}^b_{n}\right) \,,
\label{rm}
\ee
where
\be
T^{(\alpha)}_{IJ} = \fc{1}{2} \O_{KL} \S^{KL}_{IJ}  \sac
T^{(S)}_{ab} = \fc{1}{4} \O_{KL} \g^{KL}_{ab} \,.
\label{t}
\ee
This satisfies  the group property 
$R(\tilde{M}_\ione) R(\tilde{M}_\itwo) = R(\tilde{M}_\ione \tilde{M}_\itwo)$.

The vacuum energy of a system consisting of two D-branes is due 
(to the lowest order in the string coupling constant) 
to the exchange of closed strings between them, and is given by 
\be
Z =  \int_0^\infty \fc{dt}{2p^+}
\bra{B_\itwo}  e^{-(P_{cl}^- - p^-) t} \ket{B_\ione} \,,
\label{vac}
\ee
where $\ket{B_\ik}$ ($\ik=1,2$) is the boundary state for the $\ik$-th
brane, characterized by a matrix $M_\ik$. $Z$ factorizes as 
\be
Z = Z_0 \, Z_{osc} \,,
\ee
where $Z_{osc}$ depends only on the relative rotation between the branes
\be
M_{rel} \equiv M^T_\itwo \cdot M_\ione \,,
\ee
and 
\be
Z_0 = C(F_\ione) C(F_\itwo) 
\left( \tr_v \, M_{rel} - \tr_s \, M_{rel} \right) \,,
\label{zero}
\ee
where $\tr_v$ and $\tr_s$ are the traces in the vector and the spinor
representation, respectively. Note that, except for the product of the
normalization factors in $Z_0$, the vacuum energy depends only on the
relative rotation $M_{rel}$, which in turn depends only on the
relative rotation between the $SO(8)$ parts of $M_\ione$ and
$M_\itwo$, that is, on
\be
\tilde{M}_{rel} \equiv \tilde{M}^T_\itwo \cdot \tilde{M}_\ione = M_{rel} \,.
\ee


\subsection{Vacuum Energy}


As explained in the Introduction, in order to gain some understanding
about the possible presence of tachyons in the spectrum of open
strings stretched between arbitrary points of a supertube, one can
approximate the worldvolume at each of these points by that of an infinite
planar D2-brane carrying string and D0-brane charges. In this section 
we will show that the vacuum energy of two such branes vanishes exactly.

It is clear that the two D2-branes of interest will share the time 
direction, which in our Euclidean formulation we may take to be
$X^1$. They will also share one spatial direction, $X^2$ say, 
corresponding to the axis of the supertube (the $Z$-direction). 
The projection of each brane on the remaining 6-dimensional space 
orthogonal to the 12-plane is a line in this space (see Figure
\ref{projection}). For a supertube with a generic cross-section these two
lines will determine a 2-plane, the 34-plane say, and  
will form some angle $\phi$ with each other, but will not
intersect. Instead they will be separated some distance $L$ along the
remaining 4 overall orthogonal directions $X^\perp$. 
The particular case $\phi=\pi$ corresponds to a parallel 
D2/anti-D2 pair. 

\FIGURE{
\centerline
{\epsfig{file=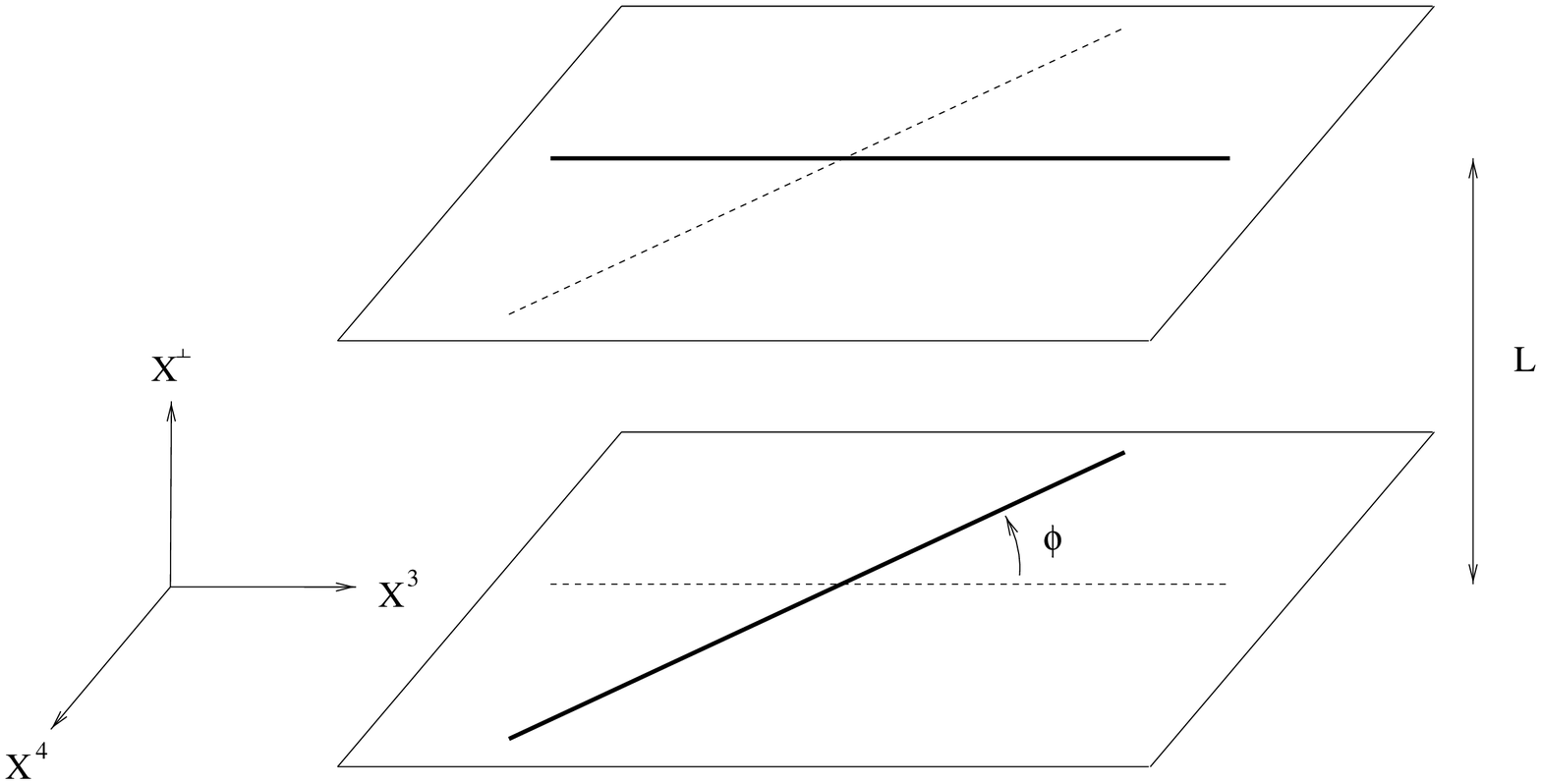, height=7cm}}
\caption{Setup of D2-branes tangent to the supertube. Both branes
share the 12-directions. The thick lines represent their projections 
on the 6-dimensional space orthogonal to the 12-plane, 
and both lie along the 34-plane. 
The `first' (`second') brane is that on the plane above (below).}
\label{projection}
}

In order to construct the boundary states describing each brane, we
assume without loss of generality that both branes are initially 
aligned along the 3-direction, and
that we then rotate the (say) second brane by an angle $\phi$. 
Thus before the rotation both branes extend along the 123-directions. 
Since they carry string charge along the 2-direction and D0-brane 
charge, the BI 2-forms $F_\ik$ ($\ik=1,2$) on their worldvolumes are
\be
F_\ik = E_\ik \, dX^1 \w dX^2 + B_\ik \, dX^2 \w dX^3 \,,
\ee
where for the moment we keep the values of the electric and
magnetic fields arbitrary. The matrices $M^\ik_{IJ}$ then take 
the form \eqn{mIJ} with
\be
M^\ik_{ij} = \fc{1}{C^2(F_\ik)} 
\pmatrix {-1 + E_\ik^2 - B_\ik^2  &     2 E_\ik      &   -2 E_\ik B_\ik  \cr
              -2 E_\ik     &  -1 + E_\ik^2 + B_\ik^2 &    2 B_\ik       \cr
           -2 E_\ik B_\ik    & -2 B_\ik   &   -1 - E_\ik^2 + B_\ik^2 \cr} \,,
\label{eqn-M-vector}
\ee
where the normalization factor is 
\be
C(F_\ik) = \sqrt{1 + E_\ik^2 + B_\ik^2} \,.
\ee
The corresponding spinorial matrix \eqn{Maa} for either 
brane is
\be
M_{ab}^\ik = \fc{1}{C(F_\ik)} \, \gamma^{12 \ldots 8} \cdot
\left( \g^{123} + E_\ik \g^3 + B_\ik \g^1 \right)_{ab} \,.
\label{eqn-M-spinor}
\ee
Note that we have kept the `plus' sign in \eqn{Maa} for both branes
because we are assuming that we start with two parallel branes of the
same type, that is, either with two D-branes or with two anti-D-branes. As
mentioned above, the D-brane/anti-D-brane case is achieved by setting 
$\phi=\pi$.

The matrix $M_\itwo(\phi)$ describing the second brane after the 
rotation is now obtained from $M_\itwo$ by conjugation with the 
rotation matrix $m(\phi)$, as in \eqn{conjugation}. 
In the present case $m(\phi)$ takes the form 
\be
\label{m}
m(\phi)_{IJ} =  \exp\left(\phi\,\S^{34}\right)_{IJ} = 
          \pmatrix { 1 &   &   &   &  \cr
                       & 1 &   &   &  \cr
                       &   & \cos \phi & - \sin \phi &          \cr
                       &   & \sin \phi &   \cos \phi &          \cr  
                       &   &           &             &  \bbi{4} \cr} 
\ee
in the vector representation, and 
\be 
m(\phi)_{ab} = \exp\left(\fc{\phi}{2}\g^{34}\right)_{ab} 
= \pmatrix{ e^{i\phi/2} \, \bbi{4} &                         \cr
			           & e^{-i\phi/2} \, \bbi{4} \cr}  
\ee
in the spinor representation. 

Now we are ready to evaluate the vacuum energy \eqn{vac} explicitly. 
The zero-mode factor \eqn{zero} is given by
\be
Z_{0}(F_\ione,F_\itwo) = \fc{\lambda}
{\sqrt{(1 + E_\ione^2 + B_\ione^2) (1 + E_\itwo^2 + B_\itwo^2)}} - 
8 \left[ (1 + E_\ione E_\itwo)\cos\phi + B_\ione B_\itwo \right] \,,
\ee
where 
\bea
\lambda &=& 2\left[ 
3 + E_\ione^2 + E_\itwo^2 + 4E_\ione E_\itwo + 3 E_\ione^2 E_\itwo^2 +
4B_\ione^2B_\itwo^2 + 2B_\ione^2(1+E_\itwo^2) + 
2B_\itwo^2(1+E_\ione^2) \right. \nn
&& \left. \hspace{1cm} + 4B_\ione B_\itwo (1+E_\ione E_\itwo)\cos\phi + 
(1+E_\ione^2)(1+E_\itwo^2)\cos 2\phi \right] \,.
\eea
In the supersymmetric case $Z_0 = 0$ and hence the vacuum energy $Z$
vanishes exactly, showing that there is no force between the branes.
To see this, recall that for a supertube with 
arbitrary shape we have $E=\pm 1$ and the only restriction on $B$ is that
it does 
not change sign along the curve. The first condition becomes 
$E_\ione= E_\itwo=\pm i$ after Euclideanization, in which case
\be
Z_0 = 8 \left( |B_\ione B_\itwo| - B_\ione B_\itwo \right) =0 \,,
\ee
which vanishes since $B_\ione$ and $B_\itwo$ have equal signs by
virtue of the second condition. Furthermore, setting 
$E_\ik = i + \ve_\ik$ and expanding $Z_0$ for small $\ve_\ik$ 
one finds that the first terms are of order $\ve^2$, 
as expected from 1/4-supersymmetry. Note that $Z_0$ might also vanish
for other values of $E_\ik, B_\ik$ and $\phi$, but we shall not
explore this possibility here.

We now turn to evaluate $Z_{osc}$.  This can be done by redefining the 
oscillators in \eqn{rm}:
\bea 
\label{diag}
\alpha_n^I \to U_{IJ}\alpha_n^J \; 
&,& \; \alpha_{-n}^I\to U_{IJ}^\dagger \alpha_{-n}^J \,, \nn 
S_n^a\to V_{ab}S_n^b\;
&,&\;S_{-n}^{a}\to V_{ab}^\dagger S_{-n}^b \,,
\eea
where the unitary matrices $U$ and $V$ are defined so that they  
diagonalize the $T$-generators \eqn{t} of the relative rotation
$M_{rel}=\tilde{M}_{rel}$, that is,
\bea
U^\dagger T^{(\a)} U &=&
\pmatrix{   i\b_1 &  &  &  &   \cr
            & -i\b_1 &  &  &   \cr
	    &  &  i\b_2 &  &   \cr
	    &  &  & -i\b_2 &   \cr
            &  &  &  &  \bbo{4}    \cr}  \,, \nn
V^\dagger T^{(S)} V &=& 
\pmatrix{ i \b_+ \bbi{2} &  &  &  \cr
          & i\b_- \bbi{2} &  &  \cr
          &  &  -i\b_+ \bbi{2} &  \cr
          &  &  &  -i\b_- \bbi{2} \cr} \,,
\eea
where 
\be
\b_\pm = \fc{\b_1 \pm \b_2}{2} \,.
\ee
Since $M_{rel}$ is effectively a rotation in the 
1234-directions it is characterized by only two angles
$\b_1$ and $\b_2$. These can be extracted from the relations 
\bea
\cos\left( \b_1/2 \right) \cos\left( \b_2/2 \right) &=& 
\fc{1}{8} \trace_s M_{rel} \nn
\cos\b_1 + \cos\b_2 &=& \fc{1}{16} 
\left[ \left(\trace_s M_{rel}\right)^2 - 2\trace_s M_{rel}^2 - 16 \right] \,.
\eea
Note that the traces are in the spinor representation, since the
vector representation of $M$ does not distinguish between brane and
anti-brane. A straightforward calculation shows that for our
supersymmetric configurations 
\be
\trace_s M_{rel} = 8 \, \mbox{sgn}(B_\ione B_\itwo) \sac 
\trace_s M_{rel}^2 = 8 \,,
\ee
and hence that $\b_1=\b_2=0$.

The evaluation of $Z_{osc}$ now proceeds as in \cite{GG96}, with the
result 
\be
Z_{osc} = \int_0^\infty  dt 
\frac{ \prod_{n=1}^\infty 
(1 - q^{2n} e^{i\b_+} )^2   (1 - q^{2n} e^{-i \b_+} )^2 
(1 - q^{2n} e^{i\b_-} )^2   (1 - q^{2n} e^{-i \b_-} )^2 }
{\prod_{n=1}^\infty  
(1 - q^{2n} )^4 (1 - q^{2n} e^{i\b_1}) (1 - q^{2n} e^{-i\b_1}) 
(1 - q^{2n} e^{i\b_2}) (1 - q^{2n} e^{-i\b_2})}\, 
\calp(t,\phi,L) \,,
\label{Zosc}
\ee
where $q=e^{-\pi t}$ and the factor 
\be
\calp(t,\phi,L) = \left\{ 
\begin{array}{lr}
V_\itwo \, (2\pi^2 t)^{-3} \, |\sin\phi |^{-1} \, 
\exp\left(-L^2/2\pi t \right) &
\mbox{\quad if $\phi\neq 0 \, \mbox{mod}\,\pi$}, \vspace{2mm} \\
V_{\it 3} \, (2\pi^2 t)^{-7/2} \, \exp\left(-L^2/2\pi t \right) \quad & 
\mbox{\quad if $\phi=0 \,\mbox{mod}\,\pi$},
\end{array} \right. 
\label{p-factor}
\ee
arises essentially 
from the delta-functions in \eqn{state} (see the Appendix). In
particular, the constants $V_\itwo$ and $V_{\it 3}$ are the (infinite)
volumes of the branes in the 12- and 123-directions respectively. 
In view of the periodicity of $Z_{osc}$, we assume that $\b_1$, $\b_2$,
$\b_+$ and $\b_-$ all lie within the interval $[0, 2\pi)$.

The result can be rewritten in a compact form in terms of Jacobi 
$\theta$-functions and the Dedekind $\eta$-function:
\be
Z_{osc} = \int_0^\infty  dt \,
\fc{\sin \left( \fc{\b_1}{2} \right) \, 
\sin \left( \fc{\b_2}{2} \right)}
{4 \sin^2 \left( \fc{\b_+}{2} \right) \, 
\sin^2 \left( \fc{\b_-}{2} \right)} \,
\fc{\theta_1^2 \left( \fc{\b_+}{2 \pi} \mid it \right) \, 
\theta_1^2 \left( \fc{\b_-}{2 \pi} \mid it \right) }
{\eta^6(it)\, \theta_1 \left( \fc{\b_1}{2 \pi} \mid it \right)
\theta_1 \left( \fc{\b_2}{2 \pi} \mid it \right)} \, 
\calp(t, \phi,L) \,.
\ee
The presence or absence of an open string tachyon is most clearly 
exhibited in the short cylinder limit, $t \to 0$. To study this we 
perform a Jacobi transformation $t \to t^\prime=1/t $. This gives 
\be
Z_{osc} = \int_0^\infty  \fc{dt'}{t'} \,
\fc{\sin \left( \fc{\b_1}{2} \right) \, 
\sin \left( \fc{\b_2}{2} \right)}
{4 \sin^2 \left( \fc{\b_+}{2} \right) \, 
\sin^2 \left( \fc{\b_-}{2} \right)} \,
\fc{\theta_1^2 \left( -i t'\fc{\b_+}{2 \pi} \mid i t' \right) \, 
\theta_1^2 \left( -i t' \fc{\b_-}{2 \pi} \mid i t' \right) }
{\eta^6(i t')\, \theta_1 \left( -i t'\fc{\b_1}{2 \pi} \mid i t' \right)
\theta_1 \left( -i t' \fc{\b_2}{2 \pi} \mid i t' \right)} \, 
\calp \left( \fc{1}{t'},\phi,L \right) \,.
\label{Z-sine}
\ee
Since the argument  of $\theta_1$ is now imaginary, the behaviour 
of the relevant part of the integrand in the short cylinder limit, 
$t^\prime\to \infty$, is given by\footnote{$Z_{osc}$ might appear to
diverge (to vanish) when $\b_1$ and/or $\b_2$ ($\b_+$ and/or $\b_-$)
vanish because of the sine factors in \eqn{Z-sine}. This is not the
case, however, as is manifest from \eqn{Zosc} or as can be seen
directly from \eqn{Z-sine} by noting that 
$\lim_{\nu \rightarrow 0} \theta_1(\nu|\tau) / \sin(\pi \nu) = 2
\eta^3(\tau)$ for any finite $\tau$. Thus the formula \eqn{integrand} gives
the correct asymptotic behaviour of the integrand for any value of
$\b_1$ and $\b_2$.} 
\be
e^{ - t' \left(L^2 - \pi|\b_1-\b_2|\right) /2\pi} + 
\calo(e^{-\pi t'}) \,.
\label{integrand}
\ee
Thus we see that generically $Z_{osc}$ diverges for separations 
$L^2 \leq \pi|\b_1-\b_2|$. These divergence signals the
appearance of a tachyonic instability. Supersymmetric configurations
are free of this instability, as we wanted to see, since in this case 
$\b_1=\b_2=0$. Note that, in fact, there is no tachyonic instability
in any brane configuration with $\b_1 = \b_2$, although generically
there will still be a force between the branes unless $Z_0=0$. Again,
we shall not explore these more general possibilities here.

\sect{Discussion}
\label{discussion}


We have generalized the 1/4-supersymmetric D2-brane supertube with
circular cross-section \cite{MT01}, to one for which the cross-section
is an {\it arbitrary} curve in $\bbe{8}$. This obviously includes 
the system of a parallel 1/4-supersymmetric D2-brane and 
anti-D2-brane of \cite{BK01} as a special case. 
In many respects, this result is counter-intuitive.
While it is not difficult to imagine how a circular tube may be
supported from collapse by angular momentum, it is less obvious how
this is possible when the cross-section is non-circular and
non-planar: one would imagine that this would imply a clash between
the need for angular momentum to prevent collapse and the need for
time-independence of the energy density to preserve supersymmetry. 
Of course this problem already arose for the elliptical supertube
found in \cite{BK01} but that was found in the context of the
Matrix model and, as we have pointed out, the implications 
of results in Matrix Theory for supertube configurations are not
immediately clear. Nevertheless, it was the results of \cite{BK01} that
motivated us to re-examine the assumption of circular symmetry for
supertubes in the DBI context and, on finding that any
cross-section is possible, to re-examine the reasons underlying the
stability of the supertube. 

We identified three sources of instability in the Introduction. The
most obvious one is the potential local instability due to D2-brane
tension and we have now dealt fairly exhaustively with that. In
particular, we have seen how the D2-brane charge in the spacetime
supersymmetry algebra can be `cancelled' by the linear momentum
generated by the BI fields; this for example allows a D2-brane extending to
infinity (and hence carrying a net charge) to have
an arbitrary cross-section while preserving supersymmetry. This simple
mechanism of `brane-charge cancellation by momentum' may have other 
applications.

A second  potential instability is due to long-range supergravity
forces. This was dealt with for a supertube with a circular cross-section in
\cite{EMT01}. Here we have generalized that analysis to an arbitrary
cross-section. A main consequence of this generalization is that the
bound on the total angular momentum of \cite{EMT01} becomes a bound on
the local linear momentum density. Violation of the bound in
\cite{EMT01} was shown to lead to the appearance of closed timelike
curves and ghost-induced instabilities on brane probes. Here we have
seen that although local violations of the bound on the momentum 
density do not necessarily imply the presence of closed timelike
curves, they do induce ghost instabilities on brane probes. 
Bound-violating solutions are consequently unstable, whereas 
bound-respecting  solutions are stable at the supergravity level.

Finally, there is a potential source of instability due to tachyon
condensation, and one of the principal purposes of this paper has been
to demonstrate that no tachyons appear in the spectrum of open strings
ending on a supertube. We have accomplished slightly less than this
since existing methods apply only to planar branes. Specifically, we
have shown that no tachyons appear in the spectrum of open strings
connecting any two tangent planes of the supertube. This includes the
case of strings connecting a D2-brane with an anti-D2-brane, provided
that the DBI fields are those required for the preservation of 1/4
supersymmetry.
\newline
\newline
{\bf Note added in proof:} While this paper was being type-written we
received \cite{BO01}, where some results which partially overlap with
those of our Section \ref{string} are obtained by different methods.

\acknowledgments

It is a pleasure to thank Michael B. Green for very helpful
discussions, and Iosif Bena for correspondence. D.M. is supported 
by a PPARC fellowship.  S.N. is supported by the British Federation 
of Women Graduates and the Cambridge Philosophical Society.

\appendix
\section{Appendix}
\renewcommand{\theequation}{A.\arabic{equation}}

In this appendix we briefly explain how the $\calp$-factor in
$Z_{osc}$ arises. For clarity, we explicitly put hats on operators.

Consider the general case of two D$p$-branes which extend along $p$
common directions (collectively denoted as $X_\|$), whose projections 
on the $34$-plane are as in Figure \ref{projection}, and 
which lie at the positions $x^\perp_\ione$ and $x^\perp_\itwo$ in 
the remaining $6-p$ overall transverse directions.
At the end we will specialize to the case of interest to us for 
which $p=2$, $X_\|=(X^1, X^2)$ and $|x^\perp_\itwo - x^\perp_\ione| = L$.

The $\calp$-factor in $Z_{osc}$ is
\be
\calp = \bra{p=0} \, \d_\itwo^\perp \, q^{\fc{\ph^2}{2}} \,
\d_\ione^\perp \, \ket{p=0} \equiv \calp_\| \, \calp_{3-4} \, \calp_\perp \,,
\ee
where $q=e^{-\pi t}$,
\bea
\calp_\| &=& \braket{\mbox{$p_\|$=0}}{\mbox{$p_\|$=$0$}} \,, \nn
\calp_{3-4} &=& \bra{\mbox{$p_3$=$p_4$=0}} \d(\xh_3') \, 
q^{\fc{\ph_3^2}{2}} \, \d(\xh_3) \ket{\mbox{$p_3$=$p_4$=0}} \,, \nn
\calp_\perp &=& \bra{\mbox{$p_\perp$=0}} \, \d(\xh^\perp - x^\perp_\itwo) 
\, q^{\fc{\ph^2_\perp}{2}} \, \d(\xh^\perp - x^\perp_\ione) 
\ket{\mbox{$p_\perp$=0}} \,,
\eea
and 
\be
\xh_3' = \cos \phi \, \xh_3 + \sin \phi \, \xh_4 \,.
\ee
The first factor is just the $p$-dimensional (infinite) volume of the
directions common to both branes, $\calp_\|=V_{\it p}$.
The second one is easily evaluated by first introducing an integral
representation of the delta function,
\bea
\calp_{3-4} &=& \int \fc{dk}{2\pi} \, \bra{\mbox{$p_3$=$p_4$=$0$}} 
\, \d(\xh_3') \, q^{\fc{\ph_3^2}{2}} \, e^{ik \xh_3} 
\ket{\mbox{$p_3$=$p_4$=$0$}} \nn
&=& \int \fc{dk}{\sqrt{2\pi}} \, q^{\fc{k^2}{2}} \, 
\bra{\mbox{$p_3$=$p_4$=$0$}} \, 
\d(\cos \phi \, \xh_3 + \sin \phi \, \xh_4) \, 
\ket{\mbox{$p_3$=$k$, $p_4$=$0$}} \,,
\eea
and then introducing a resolution of the identity in position
space,
\be
\calp_{3-4}=\int \fc{dk}{2\pi} \, q^{\fc{k^2}{2}} \, \int dx_3 \int dx_4 \, 
e^{ik x_3} \, \d(\cos \phi \, x_3 + \sin \phi \, x_4) \,.
\ee
Now we need to distinguish two cases. If $\sin \phi \neq 0$, then
integrating first over $x_4$ we get
\be
\calp_{3-4}=\fc{1}{|\sin \phi|} \int \fc{dk}{2\pi} \,  
q^{\fc{k^2}{2}} \, \int dx_3 \, e^{ik x_3} = \fc{1}{|\sin \phi|} \,.
\ee
If $\sin \phi =0$ then $|\cos \phi|=1$ and we get 
\be
\calp_{3-4}= \ell \, \int \fc{dk}{2\pi} \,  
q^{\fc{k^2}{2}} \, \int dx_3 \, e^{ik x_3} \, \d(x_3) = 
\ell \, \left( 2 \pi^2 t \right)^{-1/2} \,,
\ee
where $\ell$ is the (infinite) length of the 4-direction. 

A similar computation shows that 
\be
\calp_\perp = \left( 2 \pi^2 t \right)^{-\fc{d^\perp}{2}} \, 
\exp \left( -\fc{\left| x^\perp_\ione - x^\perp_\itwo \right|^2}
{2\pi t} \right) \,,
\ee
where $d^\perp=6-p$ is the number of overall transverse directions.

In conclusion
\be
\calp = \left\{ \begin{array}{lr}
V_{\it p} \, |\sin \phi|^{-1} \, 
\left( 2 \pi^2 t \right)^{-\fc{d^\perp}{2}} \, 
\exp \left( -\fc{\left| x^\perp_\ione - x^\perp_\itwo \right|^2}
{2\pi t} \right)  
\quad & \mbox{if $\phi \neq 0 \,\mbox{mod}\,\pi$} \,, \vspace{2mm} \\
V_{\it p+1} \, \left( 2 \pi^2 t \right)^{-\fc{d^\perp+1}{2}} \,  
\exp \left( -\fc{\left| x^\perp_\ione - x^\perp_\itwo \right|^2}
{2\pi t} \right) \quad &
\mbox{if $\phi =0 \,\mbox{mod}\,\pi$} \,, \end{array} \right.
\ee
where $V_{\it p+1} = V_{\it p} \ell$.

Specializing to the case of interest to us yields \eqn{p-factor}.



\begin{thebibliography}{80}

\bibitem{roberto}
R.\ Emparan, 
{\sl Born-Infeld Strings Tunneling to D-branes},
\plb{423}{1998}{71}, \hepth{9711106}.

\bibitem{myers}
R.\ C.\ Myers, {\sl Dielectric-Branes}, 
\jhep{12}{1999}{022}, \hepth{9910053}.

\bibitem{giant}
J.\ McGreevy, L.\ Susskind and N.\ Toumbas, 
{\sl Invasion of the Giant Gravitons from Anti-de Sitter Space},
\jhep{06}{2000}{008}, \hepth{0003075}.

\bibitem{ps}
J.\ Polchinski and M.\ J.\ Strassler,
{\sl The String Dual of a Confining Four-Dimensional Gauge Theory},
\hepth{0003136}.
 
\bibitem{enhancon}
C.\ V.\ Johnson, A.\ W.\ Peet and J.\ Polchinski,
{\sl Gauge Theory and the Excision of Repulson Singularities},
\prd{61}{2000}{086001}, \hepth{9911161}.

\bibitem{MT01}
D.\ Mateos and P.\ K.\ Townsend,
{\sl Supertubes},
\prl{87}{2001}{011602}, \hepth{0103030}.

\bibitem{BL01}
D.\ Bak and K.\ Lee,
{\sl Noncommutative Supersymmetric Tubes},
\plb{509}{2001}{168}, \hepth{0103148}.

\bibitem{savvidy}
T. Harmark and K. G. Savvidy,
{\sl Ramond-Ramond Field Radiation from Rotating Ellipsoidal
Membranes}, 
\npb{585}{2000}{567}, \hepth{0002157}.

\bibitem{BK01}
D.\ Bak and A.\ Karch,
{\sl Supersymmetric Brane-Antibrane Configurations},
\hepth{0110039}.

\bibitem{EMT01}
R.\ Emparan, D.\ Mateos and P.\ K.\ Townsend,
{\sl Supergravity Supertubes},
\jhep{07}{2001}{011}, \hepth{0106012}.

\bibitem{sen}
A. Sen, {\sl Tachyon Condensation on the Brane Antibrane System},
\jhep{08}{1998}{012}, \hepth{9805170}.


\bibitem{kappa}
E.\ Bergshoeff, R.\ Kallosh, T.\ Ort\'\i n and G.\ Papadopoulos,
{\sl Kappa-Symmetry, Supersymmetry and Intersecting Branes},
\npb{502}{1997}{149}, \hepth{9705040}.


\bibitem{dWHN}
B.\ de Wit, J.\ Hoppe and H.\ Nicolai,
{\sl On the Quantum Mechanics of Supermembranes},
\npb{305}{1988}{545}.


\bibitem{GG96}
M.\ B.\ Green and M.\ Gutperle, 
{\sl Light-cone Supersymmetry and D-branes}, 
\npb{476}{1996}{484}, \hepth{9604091}.

\bibitem{BO01}
D.\ Bak and N.\ Ohta, {\sl Supersymmetric D2 Anti-D2 Strings},
\hepth{0112034}.




\end{thebibliography}
\end{document}